# Exploring control of the emergent exciton insulator state in 1T-TiSe$_2$ monolayer by state-of-the-art theory models

Hong Tang[1,a], Li Yin[1], Gábor I. Csonka[2,1] and Adrienn Ruzsinszky[1,b]

[1] Department of Physics and Engineering Physics, Tulane University, New Orleans, LA 70118

[2] Inorganic and Analytical Chemistry Department, Budapest University of Technology and Economics, Budapest, H-1111

**ABSTRACT** The layered transition metal dichalcogenide 1T-TiSe$_2$ is of great research interest, having intriguing properties of charge density waves (CDW) and superconductivity under doping or pressurizing. The monolayer form of 1T-TiSe$_2$ also shows a CDW with a higher transition temperature T$_c$ than the bulk, indicating a stronger CDW interaction. By using the meta-generalized gradient approximation (metaGGA)-based model Bethe-Salpeter Equation (BSE) and many-body perturbation GW+BSE methods, we calculate the exciton binding energies and electron energy loss spectrum (EELS) for the 1T-TiSe$_2$ monolayer under different in-plane biaxial strains. We find that even without strain the 1T-TiSe$_2$ monolayer can have negative exciton energies at the Brillouin zone boundary point M, with a binding energy larger than the gap. The calculated EELS reinforces this picture, indicating EI (exciton insulator) states in 1T-TiSe$_2$ monolayer even without strain. The Wannier-Mott formula calculations of exciton binding energy corroborate results from GW+BSE. Small compressive strains enhance the EI state, and for tensile strains slightly less than 3%, the EI state in this monolayer persists. At large tensile strains, the material makes a transition to a normal semiconductor. Our results provide important information for understanding the quantum nature of this two-dimensional (2D) material. Our results from the standard G0W0@PBE+SOC+U+BSE approach are not qualitatively different from those of a more computationally efficient metaGGA-based SCAN+SOC+U+mBSE+$f_{xc}^{loc}$ approach that employs a model BSE.

---

[a] Email: htang5@tulane.edu
[b] Email: aruzsin@tulane.edu

Exciton insulator (EI) [1-4] is a state of matter in which electrons and holes spontaneously form bound excitons, due to strong electron-hole interactions. The formed bosonic excitons condense into Bose-Einstein condensates (BEC) at low temperatures, forming a superfluid [5], an insulating electronic crystal [4, 6], or possibly a quantum liquid crystal [7,8]. Since the prediction of EI decades ago [1, 2], the search and verification of EI candidate materials have been of great interest and are actively ongoing, mainly driven by the fascinating macroscopic quantum phenomena and applications that can be brought by EIs, such as superfluidity, perfect conductivity without a Meissner effect [8] and fractional Hall effects [9].

EI states can be realized in a bilayer quantum well under a high magnetic field [10-12], where the Landau level is vital for the EI formation and the electron and hole spatially reside in different layers, and in gate field tuned 2D moiré heterostructures [13-15], in which EI is induced by the moiré flat bands instead of the dispersionless Landau levels. Also, great attention has been paid to the naturally formed EI states in equilibrium solid state materials without any external excitations and engineered structures, such as $TmSe_{0.45}Te_{0.55}$ [16,17], $Ta_2NiSe_5$ [18], $1T-ZrTe_2$ [19,20] and $1T-TiSe_2$ [21,22]. Those materials usually feature a semiconducting or semimetallic electronic configuration with a small band gap or a small overlap between valence and conduction bands, and a relatively weak screening effect for electrons and holes.

In EIs, the pairs of bound electrons and holes with different momenta form non-zero momentum excitons, and, after Bose condensation, the inverse of the exciton momentum sets a new length scale for charge spatial modulations, usually accompanied by a structural instability at this length scale concurrently. This apparent charge density wave (CDW) effect differs little from the one caused by Peierls instability (PI) [23] or Jahn-Teller distortion [24], which is driven by the electron-phonon interaction and is unrelated to exciton formation. Bulk 1T-TiSe2 is among the intensely debated cases, regarding the EI or PI origination of the CDW phase transition observed at 200 K. The experimental investigation [25] on monolayer 1T-TiSe2 found that its CDW transition temperature is 232K, higher than that of its bulk counterpart, indicating a more robust CDW phase in the monolayer, and suggesting a possible anisotropic order parameter in the bulk. A weak-coupling BCS like exciton order parameter equation calculation [26] shows that the majority fraction of experimental $T_c$ (CDW transition temperature) is from the Coulomb interaction, supporting the EI state in 1T-TiSe2.

Strain engineering is commonly used in material engineering to effectively tune material properties. Due to their relatively high flexibility and strength, two dimensional (2D) materials can endure strains over 10% [27, 28], making properties, (such as band gap [29], spin polarization [30], magnetism [31], optical spectrum [32-34], and exciton state [35]), adjustable dramatically in 2D layer and nanoribbon forms, enabling flexible, tunable and versatile applications. In this work, we explore the strain effect on the exciton binding energy in the monolayer 1T-TiSe2 by the advanced SCAN (strongly constrained and appropriately normed) metaGGA functional [36] within the framework of linear response time-dependent density functional theory (TDDFT) and many-body perturbation methods GW [37] with BSE (Bethe-Salpeter Equation) [38]. We find that in the normal phase without strain, the binding energy of the lowest energy exciton formed by the hole at $\Gamma$ and electron at M can be 380 meV, larger than the indirect gap 98 meV between $\Gamma$ and M. The calculated EELS at different momentum transfer shows a plasmon soft mode dispersion, indicating EI states existing in this monolayer. The EI state in the 1T-TiSe2 monolayer is robust against in-plane strains, and it is enhanced even more with compressive strains, persisting up to about 3% of tensile strain. Our results provide an instrumental input to understanding and controlling the quantum properties of 1T-TiSe2.

**Theoretical and computational methods**

The standard $G_0W_0$ [37] and $G_0W_0$+BSE [38] calculations were performed in BerkeleyGW [37] by pairing with Quantum ESPRESSO [39]. The wavefunction energy cutoff is 70 Ry (~950 eV). The energy cutoff for the epsilon matrix is 18 Ry (~240 eV). The k-point mesh of $18 \times 18 \times 1$ and four valence and four conduction bands were set for optical calculations. The band number for summation is 440. The correction of the exact static remainder [40] and the slab Coulomb truncation for 2D systems were also used. The wavefunctions from PBE [41] +SOC+U (Perdew-Burke-Ernzerhof+spin-orbit-coupling+Hubbard U) were used for $G_0W_0$+BSE calculations, with U=4.45 eV. The Hubbard correction to PBE makes the bandgap of the unstrained monolayer 103 meV. After the $G_0W_0$ calculation, the gap becomes about 1 eV larger. We applied a scissor operation to the quasiparticle energies to bring the $G_0W_0$ gap close to the experimental value, keeping the $G_0W_0$ Fermi level unchanged, and performed the BSE calculation afterwards. We found that the results are not sensitive to a slight change of the $G_0W_0$ Fermi level. The Wannier-Mott formula [42] for exciton binding energy calculations is used to assess consistency of the results. We find that the SCAN+U+SOC+mBSE+ $f_{xc}^{loc}$ (see the following paragraphs) and $G_0W_0$+BSE results are approximately consistent. We note that PBE+SOC+U with U=2.5 eV, turns the unstrained monolayer to a metal, but after the $G_0W_0$ calculation, it becomes semiconducting with an indirect gap of 103 meV, close to the experimental band gap. The Wannier-Mott formula calculation of exciton binding energy for this case also leads to a result consistent with the results obtained by the methods applied in this paper.

Our TDDFT-based approximation resembles the model BSE (mBSE) method [43] in the construction of the screening. mBSE and TDDFT are both based on the Casida equation formalism. The algorithm for solving the Casida equation [44] is very similar to that for the Bethe-Salpeter equation (BSE) [38]. In the standard GW+BSE approach to optical absorption (including excitonic effects), one first computes a GW quasiparticle band structure, an expensive step scaling as $N^4$ [43] where N is the number of electrons, and then uses that to construct a frequency-dependent dielectric function with off-diagonal matrix elements. Here we propose (but only partially implement) a one-shot SCAN+U functional, with the efficiency of a semi-local metaGGA and with U fitted to yield the experimental band gap. SCAN+U replaces the originally proposed hybrid functional by A. Tal et al [43]. Because SCAN satisfies 17 exact constraints [36], while PBE satisfies only 9, and because SCAN satisfies more appropriate norms, U is typically smaller for SCAN than for PBE. At a much lower cost, SCAN+U can perform like a hybrid functional for difficult problems [45].

In our calculations, we build the mBSE screening based on the dielectric function from G0W0@SCAN+SOC+U. We did not achieve the full speed-up over GW+BSE that Ref. 43 achieved by the dielectric dependent hybrid functional [43,46], but our choice appears to be a safer technique since we did extend beyond three dimensional *sp* solids to the two dimensional transition-metal dichalcogenide 1T-TiSe$_2$, including the spin-orbit correction (SOC). We compute a static $1/\varepsilon_\infty(q)$ from G$_0$W$_0$@SCAN+SOC+U and determine the two parameters by fitting the diagonal $1/\varepsilon_\infty(q)$ to a simple model with the same two parameters. $\alpha$ and $\mu$ are obtained from a fitting of the static dielectric function with the following model [43],

$$\varepsilon^{-1}(q) = 1 - (1-\alpha)\exp\left(-\frac{q^2}{4\mu^2}\right),\qquad(1)$$

where $q$ is the length of the wavevector. The values of $\alpha$ and $\mu$ are 0.329 and 0.62, respectively, for 1T-TiSe$_2$ monolayer (see SI Figure S1).

In the standard GW+BSE approach to optical absorption, the absorption frequencies are found as generalized eigenvalues of an equation like the Casida equation of time-dependent density functional theory (TDDFT), involving a screened exchange matrix element between pairs of one-electron Bloch states. In our approach, we keep only the diagonal elements of the corresponding dielectric matrix (which is a computational saving since the diagonal of the matrix is much smaller than the whole matrix), ignore the frequency dependence, and model the effect of the off-diagonal matrix elements or local-field effects by a short-ranged static exchange-correlation kernel $f_{xc}^{loc}$, depending as in Ref. 43 on the parameters $\alpha$ and $\mu$.

We denote this method as SCAN+U+mBSE+$f_{xc}^{loc}$ where $f_{xc}^{loc}$ is a local exchange-correlation kernel built up from the SCAN meta-GGA. The local-field effects from $f_{xc}^{loc}$ were found to have minor impact in bulk solids [43] but they could be more important in low-dimensional materials. MetaGGA-based kernels are a promising direction for excitonic effects [47-50].

The calculations were conducted in the Vienna Ab initio Software Package (VASP 5.4.4) [51] with projector augmented-wave (PAW) pseudopotentials [52]. The Perdew-Burke-Ernzerhof (PBE) [41] functional with spin-orbit coupling (SOC) relaxed monolayer 1T-TiSe2 has an in-plane lattice constant $a$=3.528 Å, while the SCAN result is 3.541 Å, and both are close to the experimental value of 3.538 Å [53]. In the following calculations, for the unstrained monolayer, we use the experimental value $a$=3.538 Å and fully relaxed atomic coordinates. For the strained monolayers, $a$ is adjusted accordingly, corresponding to biaxial compressive 0.25%, biaxial tensile 2.0% and 3.0% strains, and the atomic coordinates are fully relaxed to all forces less than 0.008 eV/Å. The monolayer system was simulated with a 16 Å vacuum region to suppress interactions between adjacent layers. A plane wave energy cutoff of 450 eV and an $18 \times 18 \times 1$ k-point mesh were used for the calculations. The SCAN+SOC+U (U=3.81 eV) method is used for ground state calculations and the calculated indirect gap between Γ and M for the unstrained case is 100 meV, close to the experimental value of 98 meV at room temperature [25].

**Results and discussions**

**Strain effects on exciton binding energies** The structure of the 1T-TiSe2 monolayer is shown in Figure 1a. A Ti atom and its six adjacent Se atoms form a slant octahedron (or a trigonal antiprism). Those octahedra form an edge-sharing network in the 2D plane. Figures 1b and 1c show the partial density of states (PDOS) resolved band structure of the unstrained 1T-TiSe2 monolayer. The conduction bands arise mainly from Ti 3d orbitals while the valence bands around the Fermi level are from Se 4p orbitals, consistent with previous calculations [26], and with the results of bulk 1T-TiSe2 [25, 54]. Figures 1d-1f show the spin projections of band states in different spatial directions. The spin of the top of the valence band around Γ, contributed from the Se-p orbital, is mainly oriented along the z axis, while the conduction band minimum around M, contributed from the Ti-d orbital, is mainly oriented along the x-y plane. The z axis is along the supercell vector c, the x axis is in the monolayer plane and along the equally bisecting line for the angle formed by the cell vectors a and b, and the y axis is also in the monolayer plane. Figures 2a-d show the quasiparticle and the first (lowest energy) exciton band structures along the Γ-M direction of the monolayer, under different strains, calculated from $G_0W_0$(@ PBE+SOC+U 4.45 eV)+BSE. For compressive 0.25%, zero, and 2% tensile strains, we find a portion of the exciton bands below VBM, indicating negative exciton energies and hence larger exciton binding energies than the quasiparticle gaps at M. The excitons can be formed spontaneously under these strain conditions, without any excitation energies, indicating that the EI state is the ground state. For 3% tensile strain, the exciton band is all above the VBM, and a non-zero excitation energy is needed to excite and form excitons, showing a normal semiconducting state. Figures 2e and 2f show the changing trend of the quasiparticle gaps and the first exciton binding energies at M with strains. Figure 2e is the results from $G_0W_0$(@PBE+SOC+U)+BSE (U=4.45 eV) and Figure 2f shows results from SCAN+SOC+U+mBSE+$f_{xc}^{loc}$ (U=3.81 eV). They both show almost linear dependences on the gap and exciton binding energy with the applied strains from compressive 0.25% to tensile 4%. The $G_0W_0$(@PBE+SOC+U)+BSE method predicts that below the tensile strain of 2.75%, the binding energy of the exciton is larger than the gap. Beyond that strain, the gap becomes larger, indicating an EI state below strains of 2.75%, and a normal semiconducting state above strains of 2.75%. The calculated exciton binding energy from SCAN+SOC+U+mBSE+$f_{xc}^{loc}$ is about 40 meV lower than that from $G_0W_0$(@PBE+SOC+U)+BSE, making the EI to semiconducting transition strain value slightly smaller (about tensile 2.66%). The calculated relatively large binding energy (~360-460 meV) of the exciton formed by electron and hole with a momentum transfer from $\Gamma$ to M indicates a strong interaction between electrons and holes in the 1T-TiSe2 monolayer, consistent with the reduced screening effect from the reduced dimensionality of the 2D layer.

The exciton binding energy can be estimated by the Wannier-Mott formula:

$$E_b = -\left(m_\mu * Ry\right)/\varepsilon_r^2 \ , \tag{2}$$

where $m_\mu = (m_e \cdot m_h)/(m_e + m_h)$ is the reduced mass, $m_e$ and $m_h$ are the effective masses (referenced to the rest mass of a bare electron) of the electron at the conduction band minimum and the hole at the valence band maximum, respectively. $Ry$=13.6 eV and $\varepsilon_r$ is the dielectric constant of the system as referred to vacuum, at the wavevector of the exciton. The calculated binding energy of the first exciton with momentum transfer from $\Gamma$ to M is shown in Table 1. The effective masses of the exciton estimated from the $G_0W_0$ ($G_0W_0$(@PBE+SOC+U)) band structure are slightly larger than those from the DFT (PBE+SOC+U), resulting in slightly larger binding energies from the former (about 100 meV larger) than the latter. The estimated exciton binding energies from the Wannier Mott model are close to the results from standard $G_0W_0$(@PBE+SOC+U)+BSE and the new SCAN+SOC+U+mBSE+$f_{xc}^{loc}$, as shown in Figure 2;

all fall in the ranges of ~350-520 meV, larger than the band gap (~100meV) of the unstrained monolayer. For $G_0W_0$(@PBE+SOC+U), with U=2.5 eV, the result of PBE+SOC+U is a metallic state for the unstrained 1T-TiSe2 monolayer with the overlap of about 310 meV between the VBM at Γ and CBM at M. After the $G_0W_0$ calculation, the material becomes semiconducting with an indirect gap of 103 meV between Γ and M. By using the Wannier function interpolation [55] of the calculated $G_0W_0$ band structure, the masses of the electron at M and hole at Γ can be determined. With the Wannier-Mott formula, the exciton binding energy with the momentum transfer from Γ to M is calculated as 303 meV, for $G_0W_0$@PBE+SOC+U=2.5 eV, still close to the values determined by the other methods considered in this work. This approximate consistency of the calculated exciton binding energies with different methods confirms the strong interaction between electron and hole and the EI states in this monolayer.

With the strain altering from compressive 0.25% to tensile 3%, the band gap between Γ and M is increasing, opposite to the trend from monolayer 1H-MoS2 with strain [56]. This increasing gap trend can be understood as the change of the separation between the occupied bonding $\sigma$ and unoccupied antibonding $\sigma^*$ bands, and the width of the unoccupied non-bonding $d$-orbital bands [57], resulting from the strain-induced change of the ligand field. With strain increasing from compressive to tensile, in the octahedron formed by one Ti and six Se atoms, the adjacent Ti-Se bonds maintain almost perpendicular orientation with a slight deformation. The Ti-Se bond length increases (see SI Figure S9 and Table S1), resulting in reduced strength of the ligand field, and hence the decrease in the d-orbital band splitting and band width. The total effect is an increase of the Γ − M indirect band gap formed between the top of bonding $\sigma$ and the bottom of the d bands within the $\sigma$-$\sigma^*$ gap (see SI Figure S10 and Table S2). Compared with the large change of the band gap of about 600% from compressive 0.25% to tensile 3% strain, the exciton binding energy change is relatively small, only about 10%, indicating a small change in the effective mass of the exciton, the screening effect in the monolayer, and the robust exciton states.

**Electron energy loss (EELS) and plasmon dispersion**

Figure 3 shows the calculated EELS [58] curves of different momentum transfer for different strain conditions of the 1T TiSe2 monolayer. The onset frequency of each EELS curve at different momentum transfer (q1, q2, …), indicated by the dashed black lines, represents the starting point of the energy loss due to creations of excitons with that momentum. The curve q1 represents zero momentum transfer of electron and hole in the exciton formation, q2 for a momentum transfer of $0.1111Q_{\mathrm{M}}$, where $Q_{\mathrm{M}}$ is the length from $\Gamma$ to M in BZ, q3 for $0.2222Q_{\mathrm{M}}$, q4 for $0.3333Q_{\mathrm{M}}$, …, and q10 for $1.0Q_{\mathrm{M}}$. The positive EELS at the onset frequencies, in the curves q1-q5 of figure 3a, corresponds to energy inputs needed to excite excitons and plasmons with those momenta. The EELS curves q6-q10 show anomalous features around zero frequency in figure 3a, indicating that one may gain energy for exciting excitons and plasmons with those momenta. This anomaly reflects that the binding energy of excitons is already larger than the gap, and no energy input is needed to excite excitons. The EI state can spontaneously form in the material. The EELS curves in figure 3b are approximately like those in 3a, due to the small compressive strain in figure 3a. The onset frequencies of curves q1-q5 from the unstrained monolayer are larger than those of the compressed one. The curves of q6-q10 also show an anomaly around zero frequency for the unstrained monolayer, indicating that the EI state can also spontaneously form. With an increase of tensile strain to 2% in figure 3c, the onset frequencies of curves q1-q5 are even increased, and the EELS curves of q6 and q7 become positive at the onset frequencies. Only the curves of q8-q10 show a soft plasmon around zero frequency. This reflects the fact that the increase of the gap induced by tensile strains is faster than the increase of exciton binding energy, as shown in figure 2, so, the range of q points with negative exciton energies is shrinking with tensile strains, as shown in figure 2c. At the tensile strain of 2%, the monolayer can also form the EI state. With the tensile strain increased to 3%, (figure 3d), all the EELS curves turn to positive at the onset frequencies around zero frequency, although the onset frequencies for curves q9 and q10 are small. This means an unspontaneous EI state. For exciting excitons with all the momenta, an energy input is needed, hence the monolayer is a normal semiconductor, and the EI state cannot be formed. One can observe strain induced EI-to-semiconducting phase transition in this monolayer.

The onset frequencies of the calculated EELS curves with small momentum transfers (q1-q3) of the unstrained monolayer in Figure 3b show an increasing trend with small momentum transfer q values. This is qualitatively consistent with experiments of the bulk 1T TiSe2 [21, 59], where at the normal phase, the low energy plasmon frequency increases with q at small q values, while with increasing q, the low energy plasmon peak is quickly undiscernible due to the strong decay induced by the Landau damping [60]. Monolayer 1T TiSe2 is slightly different from its bulk counterpart. Due to the strong quantum confinement effect, the Ti d orbitals repel the Se p orbitals, and form a relatively larger gap (~100 meV) in the monolayer, while in the bulk, TiSe2 either has a smaller gap or is semimetal with both electron and hole pockets around the Fermi level in the normal phase [61]. Possible interband transitions from different layers in the bulk case are absent in the monolayer. This will lead to quantitatively different low energy plasmon frequencies and possibly different trends with q.

Figures 4a-4f show the exciton spectra for different momentum transfer of the unstrained 1T-TiSe2 monolayer. The first (lowest energy) excitons are all dark excitons, while for small momentum transfer (see figures 4a and 4b), there are bright excitons with energies very close to the first dark excitons. For this small momentum transfer region, these almost degenerate bright

and dark excitons may be useful for manipulating quantum information. The formed lowest energy dark excitons reflect the fact that the hole from the top of valence band has spin mainly polarized in the z direction, while the spins of the electrons in the lowest conduction band are mainly polarized in the x-y plane. The relatively large lifetime of these dark excitons may also be a useful tool for important quantum manipulations. Furthermore, Figures 4(g)-4(l) show that those dark excitons have a very concentrated distribution in the momentum space. Figures 4(m)-4(n) show the modulus squared of the real-space exciton wavefunction of the lowest energy exciton at momentum transfer q1=0, showing a relatively large spatial distribution and a circular shape with a radius of about 1-2 nm. Figures 4(o)-4(p) show the spatial distribution of the exciton wave function of the lowest energy exciton at q10=$1.0Q_M$, showing a delocalization feature in real-space. This wavefunction delocalization feature also occurs in monolayer transition metal dichalcogenide topological exciton insulators [62].

**Conclusions**

In conclusion, we calculated the exciton binding energies and EELS for the 1T-TiSe monolayer under different in-plane strain conditions with a SCAN metaGGA based model BSE with local $f_{xc}$, and many-body perturbation GW+BSE methods. SCAN+U+mBSE+$f_{xc}^{loc}$ appears as a promising and computationally more feasible alternative to GW+BSE. We found that even without strains, the 1T-TiSe$_2$ monolayer can have negative exciton energies at the BZ boundary point M. Those excitons with larger binding energy than the gap can be formed spontaneously without any external energy input. The calculated EELS also reinforces this picture, indicating that the 1T-TiSe$_2$ monolayer is an EI even without any strain. The exciton binding energy calculations based on the Wannier-Mott formula also deliver consistent results. Small compressive strains enhance this EI picture for this monolayer, and even for tensile strains of about a little bit less than 3%, the EI state in this monolayer persists. At large tensile strains, the 1T-TiSe monolayer transitions to a normal semiconductor. The gap tunability by strain can be ~120 meV/% in 1T TiSe2 monolayer, indicating a promising application in novel electronic devices with high flexibility. Also, our results provide important information for understanding the quantum nature of this amazing 2D material.

Table 1. The calculated exciton binding energy $E_b$ (in meV) from Eq. (2) of the first (lowest energy) exciton with momentum transfer from $\Gamma$ to M of 1T-TiSe2 monolayer under different strain conditions. $m_e$ and $m_h$ are the effective masses of the electron at the conduction band minimum and the hole at the valence band maximum, respectively. $m_\mu$ is the reduced mass of the exciton. All masses are referenced to the rest mass of a bare electron. $\varepsilon_r$ is the dielectric constant of the system as referred to vacuum, at the wavevector q from $\Gamma$ to M, calculated using $G_0W_0$@PBE+SOC+U. The PBE+SOC+U block is the result from the masses of electron and hole calculated from the PBE+SOC+U band structure, while the $G_0W_0$ block is from that of the $G_0W_0$ band structure. The biaxial in-plane strains are applied with a negative sign for compressive strain and a positive sign for tensile.

| | PBE+SOC+U | | | | $G_0W_0$@PBE+SOC+U | | | |
|---|---|---|---|---|---|---|---|---|
| Strains(%) | -0.25 | 0.0 | 2.0 | 3.0 | -0.25 | 0.0 | 2.0 | 3.0 |
| $m_e$ | 1.003 | 0.998 | 1.144 | 1.235 | 1.013 | 1.089 | 1.107 | 1.154 |
| $m_h$ | 0.143 | 0.144 | 0.153 | 0.159 | 0.212 | 0.209 | 0.194 | 0.194 |
| $m_\mu$ | 0.125 | 0.125 | 0.135 | 0.141 | 0.175 | 0.175 | 0.165 | 0.166 |
| $1/\varepsilon_r^2$ | 0.219 | 0.219 | 0.217 | 0.216 | 0.219 | 0.219 | 0.217 | 0.216 |
| $E_b$ | 372.0 | 373.4 | 399.6 | 415.0 | 521.1 | 520.7 | 489.0 | 488.7 |

Figures

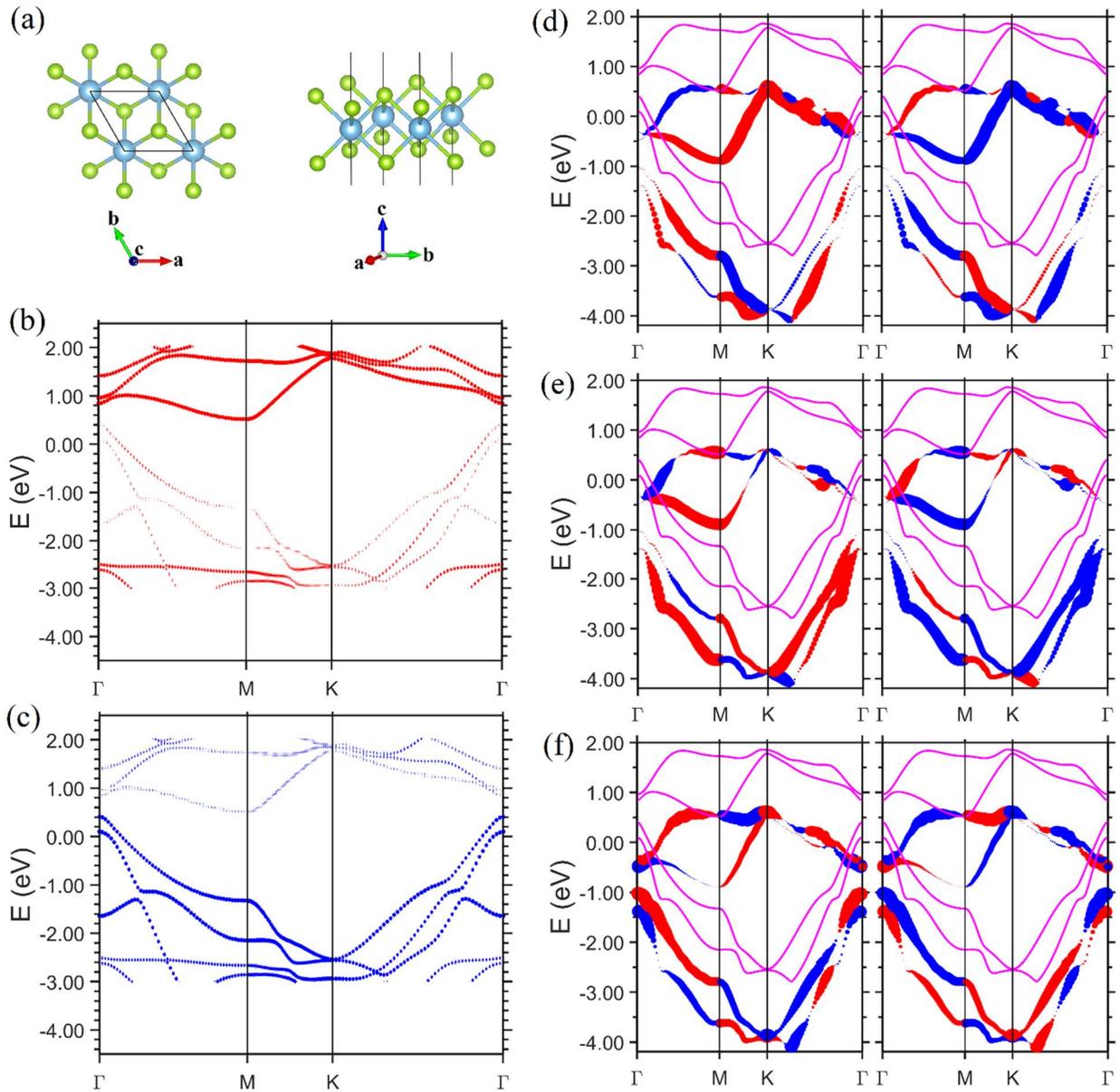

Figure 1. The geometric structures, partial density of states (PDOS) resolved band structures and spin polarization resolved band structures of the unstrained 1T-TiSe2 monolayer. (a) 1T phase TiSe2 monolayer structures in two views with blue balls for Ti atoms and green ones for Se. The black lines indicate the supercell. (b) Ti 3d orbital and (c) Se 4p orbital resolved band structures. Results are from calculations of PBE+SOC+U 4.45 eV. (d) Spin polarization resolved band

structures with spin projection on the x axis, (e) spin projection on the y axis and (f) on the z axis. The z axis is along the supercell vector c, the x axis is in the monolayer plane and along the equally bisecting line for the angle formed by the cell vectors a and b, and the y axis is also in the monolayer plane. Magenta curves are band structures from PBE+SOC+U 4.45 eV, and blue and red colored curves are from $G_0W_0@$ PBE+SOC+U 4.45 eV. The size of the blue and red dots is proportional to the magnitude of the spin projection. Red represents up-spin and blue for sown-spin. The scissor operation is done for $G_0W_0$ bands to bring the $G_0W_0$ gap close to the experimental value. In plots (d)-(f), spin polarization resolved bands with odd number indexes are shown in the left panel and the even number indexed bands are shown in the right panel for clarity, since some bands are almost degenerate in energy. The energy zero for the band structure of each approach is arbitrary.

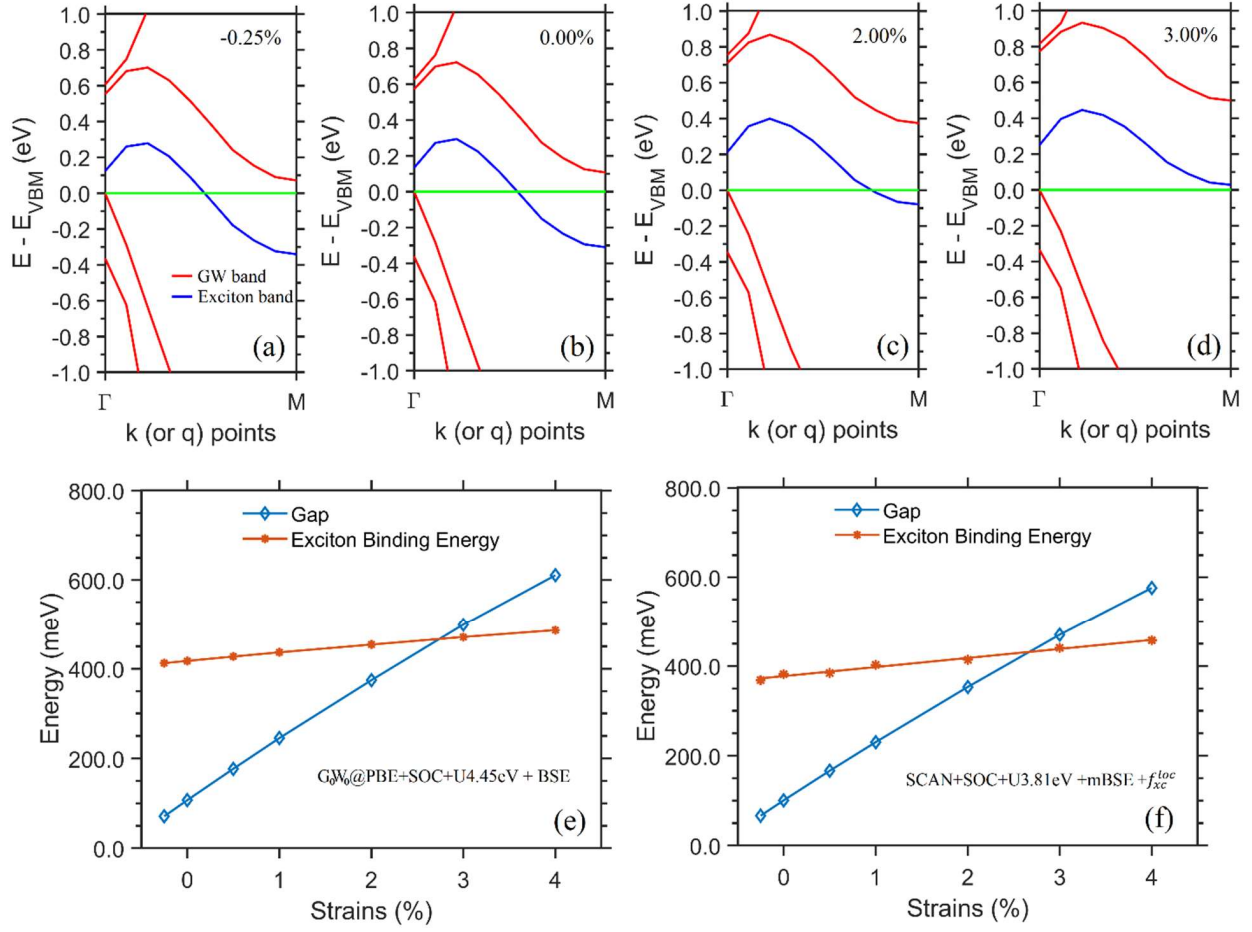

Figure 2. Top row: the quasiparticle and the first (lowest energy) exciton band structures obtained from $G_0W_0$(@ PBE+SOC+U 4.45 eV)+BSE method along the $\Gamma$ - M direction of 1T-TiSe2 monolayer under different biaxal in-plane strains. (a) compressive strain of 0.25%, (b) no strains, (c) tensile 2% and (d) tensile 3%. The energy of the valence band maximum is set to zero. Bottom row: the quasiparticle gaps and the first exciton binding energies at M in the Brillouin zone for the 1T-TiSe2 monolayer as a function of strains. (e) results from $G_0W_0$(@ PBE+SOC+U 4.45 eV)+BSE and (f) from SCAN+SOC+U (U=3.81 eV) +mBSE+$f_{xc}^{loc}$.

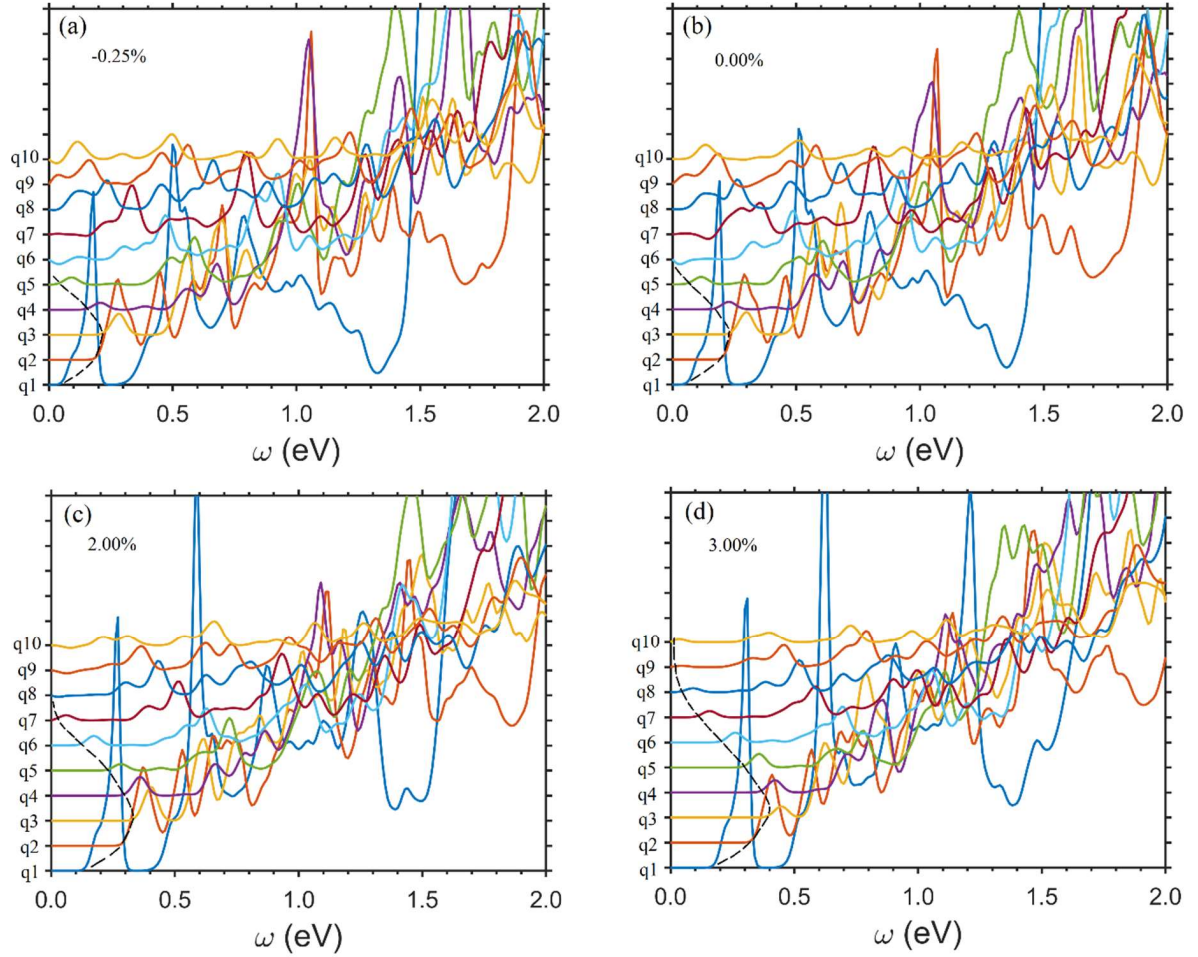

Figure 3. The calculated EELS curves of different momentum transfer of the 1T-TiSe2 monolayer under different strain conditions. The onset frequencies of the curves are connected by the dashed black lines to show the soft plasmon dispersion trend for each strain. (a) is for a compressive strain of 0.25%, (b) unstrained, (c) tensile 2% and (d) tensile 3%. The curve q1 represents zero momentum transfer of electron and hole in the exciton formation, q2 for a momentum transfer of $0.1111Q_M$, where $Q_M$ is the length from Γ to M in BZ, q3 for $0.2222Q_M$, q4 for $0.3333Q_M$, …, and q10 for $1.0Q_M$, where $Q_M$ is the wavevector from Γ to M.

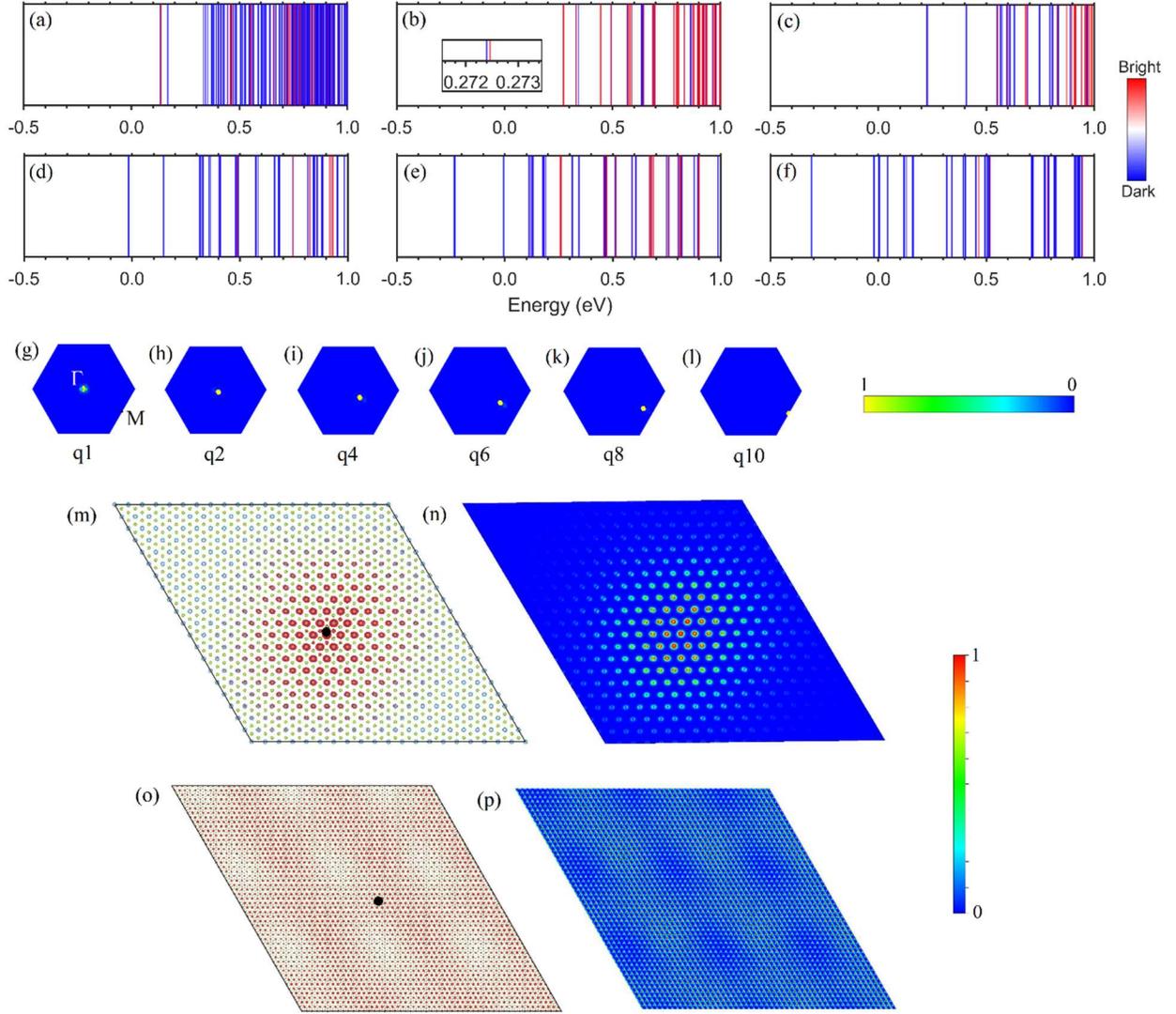

Figure 4. Rows one and two: the exciton spectra of the unstrained monolayer at different momentum transfers with (a) for q1=0, (b) q2=0.1111$Q_M$, (c) q4=0.3333$Q_M$, (d) q6=0.5556$Q_M$, (e) q8=0.7778$Q_M$, and (f) q10=1.0$Q_M$, where $Q_M$ is the length from Γ to M in BZ. Red for bright excitons and blue for dark ones. The inset in (b) shows the enlarged view of the bright and dark excitons around 0.272 eV. Row three: the amplitude of the first (lowest energy) exciton wave function in momentum space with (g) for q1=0, (h) for q2, (i) for q4, (j) for q6, (k) for q8 and (l) for q10. The hexagons represent the first BZ. The color bar in the third row represents the amplitude scale. Row four: the left plot is the modulus squared of the real-space exciton wavefunction of the lowest energy exciton at momentum transfer q1=0. The right plot shows the profile of the modulus squared of the real-space wavefunction of the same exciton in a plane close to the Ti atom layer. Row five: similar plotted and arranged plots for the lowest energy exciton at q10=1.0$Q_M$. In Rows four and five, the hole positions are denoted as black dots in the left plots and the vertical color bar represents the scale for the right two plots. All plots are for the unstrained 1T-TiSe2 monolayer.